\documentstyle[11pt,newpasp,twoside,epsf,psfig]{article}
\index{instructions}
\index{guidelines}

\def\edcomment#1{\iffalse\marginpar{\raggedright\sl#1\/}\else\relax\fi}
\reversemarginpar

\begin{document}
\title{A Rocket Model of Neutrino Jet for Pulsar Kick}
 \author{Q.~H.~Peng, L.~D.~Zhang \& X.~L.~Luo}
\affil{TDepartment of Astronomy, Nanjing University, Nanjing
210093, China Email: qhpeng@nju.edu.cn}
\author{C.~K.~Chou}
\affil{Institute of Astronomy and Department of Physics, National
Central University, Chung-Li Taiwan 32054, China Taipei}

\begin{abstract}
On the basis of the neutrino emission from the isotropic $^1$S$_0$
neutron superfluid vortexes in neutron star interiors, we propose
a rocket model of neutrino jet for the observed pulsar kick.
\end{abstract}

\section{Introduction}

Ever since the discovery of pulsars in 1967, astronomers are very
surprised to discover that the space velocities of the pulsars are
generally rather too high. Except these recycle pulsars just
mentioned, of the total 94 pulsars, the percentage of the number
of Single-pulsars with velocities exceed 100 km s$^{-1}$, 200 km
s$^{-1}$, 300 km s$^{-1}$, 500 km s$^{-1}$ and 1000 km s$^{-1}$
are respectively $79 \%$, $67\%$, $40\%$, $16\%$ and $5.6\%$. This
means that the high-velocity pulsars are rather common phenomena
in interstellar space and the cosmos. The space velocities of
these pulsars are far above both their progenitor stars and normal
stars (about 20-40 km s$^{-1}$).

At the present stage, most of the theories conceived by
astronomers to explain the unanticipated high velocities of the
nascent pulsars have been established on the basis of certain
spatially asymmetric cause due to fundamental physics or dynamical
effect during the very short interval of supernovae explosion to
form pulsars or neutron stars. Until now, four different kinds of
mechanisms for the observed neutron star kicks have been
proposed(Arras \& Lai, 1999; Kusenko \& Segre, 1999; Kusenko,
1999; Lai \& Qian, 1998a and 1998b; Lai, 2001; Lai , Chernoff \&
Cordes, 2001).

It seems that none of the four different kinds of mechanisms could
convincingly and successfully explains the observed huge pulsar
kicks. We therefore will present the idea of a gradual
acceleration due to neutrino jet rocker model for the pulsar kicks
in the talk.

\section{A Rocket Model of Neutrino Jet for the Pulsar Kick}

In 1982, Peng and his collaborators (Peng, Huang \& Huang, 1982)
proposed that neutrino cyclotron emission from the superfluid
vortex neutrons may be used as a possible mechanism for pulsar
spin down. Starting from this theory and making use of the spatial
asymmetry of neutrino spin due to parity nonconservation in weak
interaction, we now propose a new mechanism for pulsar kicks based
on a neutrino rocket model from the superfluid vortex neutrons.
The main idea is as follows:

We note that neutrons in circular motion can emit a
neutrino-antineutrino pair via the neutral current in the unified
theory of electro-weak interaction. Similarly, the super fluid
vortex neutrons can also emit neutrinos and antineutrinos (Peng,
Huang \& Huang, 1982). The neutrons in both of these processes
will lose their energy $E_{n}^{(rot)}$ and angular momentum
$\vec{J}_{n}$.

\begin{figure}
\centerline{\psfig{file=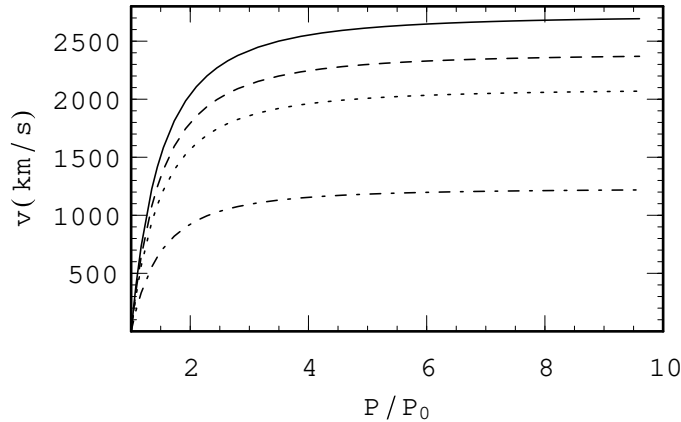,width=9 cm}} \caption{The
acceleration curves for the neutrino jet rocket model without
magnetic field decay. The spatial kick velocity of pulsars is
plotted as a function of the dimensionless variable $P/P_{0}$
where $P$ and $P_{0}$ respectively denotes the period and the
initial period. The curves (from top to bottom) corresponds
respectively to $P_{0}=0.4 \;{\rm ms}$, $0.45\;{\rm ms}$,
$0.5\;{\rm ms}$ and $0.7 \;{\rm ms}$. The parameter $ \xi =1.0
\times 10^{17}$ here.}
\end{figure}

It is expected that the angular distributions of the neutrinos and
antineutrinos are not asymmetrical due to the angular momentum
loss from the neutrons, although we have $N(\nu)=N(\bar{\nu})$
according to lepton conservation. More specifically, if the
angular momentum of the neutron is carried away primarily by the
neutrinos, the direction of the emitted neutrinos with left hand
helicity must be opposite to the original neutron angular
momentum, while the neutron itself will receive a recoil along its
angular momentum. Fortunately, it is shown by the very recent
observational evidence that the kicks of the two youngest pulsars
(the Crab pulsar and the Vela pulsar) are basically consistent
with their spinning axes (Lai, Chernoff \& Cordes, 2001). Hence,
it is anticipated that the angular momentum of the superfluid
vortex neutrons is mainly carried away by the neutrinos rather
than by antineutrinos.

\section{Pulsar Kick generated by Neutrino Rocket Propulsion}

The neutrino luminosity due to the superfluid vortex neutrons in
the neutron star interior has been derived by Peng et al. Making
use of their result, we may study the acceleration of the nascent
and determine the kick velocity $\vec{v}$ of the neutron star with
mass $M$ in terms of the rocket propulsion model. The recoil
acceleration of the neutron star due to the effect of neutrino
rocket propulsion as briefly discussed above may be written as $M
\frac{d\vec{v}}{dt}=\eta_{\nu} \frac{W_{\nu}}{c}$, here $W_{\nu}$
is the power for neutrino emission from the superfluid vortex
neutrons in the neutron star. $M$ is the mass of the neutron star.
The effective asymmetry coefficient $\eta_{\nu}$ for the neutrino
emission maybe estimated to be $\eta_{\nu} \approx 5-20$$\%$. It
is found (Peng, Huang \& Huang, 1982) that $W_{\nu} =
bG(n)\Omega$, where $\Omega$ represents the rotating angular
velocity of the neutron star. $b \cong 4.61 \times 10^{22}
R_{6}^{3}$ ( in c.g.s. unit), $R_{6}$ is the radius of the neutron
star in unit of $10^{6}$ km. $G(n) =
\frac{\overline{n^{7}}}{\overline{n}}$, where $n$ denotes the
quantum number of the neutron superfluid vortexes and it decreases
when the pulsar period increases during the dynamical evolution of
pulsars spin down. Making a working assumption $G(n) = G(n_{0})
(P/P_{0})^{-\beta}$ ($\beta \approx 3$), the corresponding recoil
velocity of the neutron star may then be obtained by using the
spin down law for the pulsars in our hybrid model. The results for
the neutron star kicks with different initial periods are shown in
Fig.~1 and Fig.~2. In these figures, the pulsar velocity $v$ is
shown as a function of $x = P/P_{0}$ (more precisely
$v(P/P_{0})-v_{0}$ ).

\begin{figure}
\centerline{\psfig{file=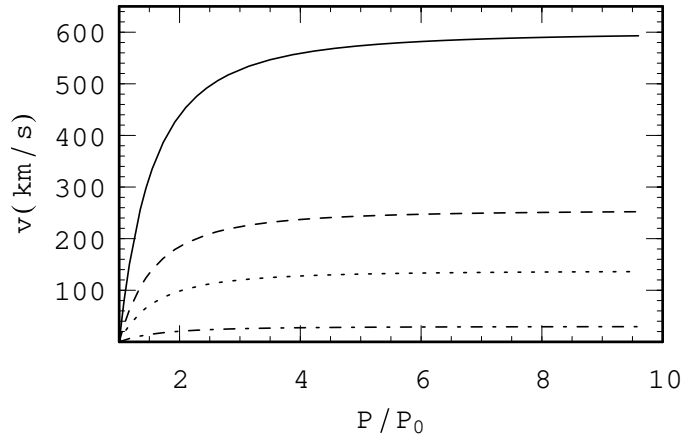,width=9 cm}} \caption{The
comparison of the calculated pulsar velocity with observation in
the absence of magnetic field decay. The pulsar period $P$ is
treated as the independent variable here. All the curves computed
from our model ( from top to down ) corresponds respectively to
$P_{0}=1\; {\rm ms}$, $1.5\; {\rm ms}$, $2\; {\rm ms}$ and $4\;
{\rm ms}$. Here the magnetic fields is $B_{12}=10^{12}$ gauss.}
\end{figure}

We will now elaborate on some of the important and subtle points
of our model. We note that the initial neutron superfluid vortex
quantum number $n_{0}$ and the initial spinning period $P_{0}$ are
the two most important parameters of our model. In order the
neutron star to receive larger pulsar kicks of several hundred km
s$^{-1}$ or even exceed 100 km s$^{-1}$ from the acceleration
scenario predicated by our neutrino jet rocket mechanism, it is
required that the initial vortex quantum number $n_{0}$ reach
(500-700) for the model without magnetic field decay, or $n_{0}=$(
5-8)$\times 10^{3}$ for models with magnetic field decay. Such
huge initial vortex quantum numbers are possible vis a vis the
chaotic and violent process during which the neutron stars are
born. This is because the spinning angular velocity $\Omega$ of
the neutron star becomes much faster after the collapse due to the
conservation of the angular momentum of the entire star. We easily
derive the result $\Omega_{0} \geq 1.0 \times 10^{4}$s$^{-1}$,
$P_{0}<1$ ms. However, for a rapidly rotating stable neutron star,
a considerable amount of the spinning angular momentum must have
been converted into the highly chaotic and turbulent (classical)
whirlpool vortexes. Moreover, it is also expected that the
turbulent vortexes can be further converted into quantized
superfluid vortexes with very high initial vortex quantum number
$n_{0}>10^{4}$ or more provided that the temperature of the
neutron star decreased down to $T<T_{trans}=2 \times 10^{10}$ K.

\section{Main Results in Our Model}
It is shown that our theory predicts naturally the gradual
acceleration of the nascent pulsars during the early stage from
$P_{0}$ to $10P_{0}$ (about 200-300 years, where $P_{0}$ denotes
the initial spinning period) and that huge natal kicks of neutron
stars exceed 1000 km s$^{-1}$ follows very nicely from our model.
We have investigated the acceleration scenario during the early
stage of pulsar evolution in terms of the initial periods and the
initial magnetic fluid. In particular, (a) the observed alignment
of the pulsar kicks with their spinning axes may be interpreted
naturally. (b) All high velocity pulsars with spatial velocities
higher than 100 km s$^{-1}$ have initial periods shorter than
(2-3) ms; moreover, the initial periods of these pulsar with huge
kicks exceed 1000 km s$^{-1}$ are shorter than 0.8 ms. (c) For the
same magnetic fields, the initial periods of the high velocity
pulsars are short. (d) For the same initial spinning pulsar
periods, the high velocity pulsars have weaker magnetic field.

\acknowledgments
This research is supported by the National
Natural Science Foundation of China (No. 10173005, 10273006 and
19935030) and a grant from the National Education Ministry of
China for Ph.D candidate training(2000028417).

\end{document}